\documentclass{zapiski}
\originfo{...}{}{}{2026}
\date{1 March 2026}

\usepackage[T1, T2A]{fontenc}
\usepackage[russian, english]{babel}

\usepackage{url}
\usepackage{latexsym}
\usepackage{multirow}
\usepackage{graphicx}
\usepackage{amssymb}
\usepackage{amsmath}
\usepackage{hyperref}
\usepackage{booktabs}
\usepackage{tabularx}
\usepackage{caption}
\usepackage{subcaption}
\usepackage{cite}
\usepackage{paralist}
\usepackage{xcolor,colortbl}
\usepackage{array}
\usepackage{setspace}
\usepackage{placeins}
\usepackage{float}
\usepackage{tikz}
\usepackage{listings}
\usepackage{siunitx}
\restylefloat{table}

\usetikzlibrary{positioning,arrows.meta}

\lstdefinestyle{promptstyle}{
  basicstyle=\ttfamily\small,
  breaklines=true,
  columns=fullflexible,
  frame=single,
  keepspaces=true,
  showstringspaces=false
}

\english

\begin{document}
	
\title[Improving Small LLMs for Code Generation with RLVR]{Improving Small Language Models for Code Generation with Reinforcement Learning from Verification Feedback}

	\author[E. Skopin]{E. Skopin}
	\address[E. Skopin]{Vyatka State University}
	\email{eclipsingstasr00@gmail.com}
	
	\author[E. Kotelnikov]{E. Kotelnikov}
	\address[E. Kotelnikov]{European University at St. Petersburg}
	\email{kotelnikov.ev@gmail.com}

    \begin{abstract}
    Reinforcement learning with verifiable rewards (RLVR) trains language models using programmatically checkable signals such as unit-test outcomes, enabling direct optimization for functional correctness in code generation.
    We conduct an empirical study of RLVR for Python code generation on the MBPP benchmark using two small models (Qwen3-0.6B and Llama3.2-1B) with LoRA fine-tuning.
    Across multiple reward formulations such as: unit-test-only rewards, static-analysis-only shaping via the Ruff linter, and a combined reward, we compare group-based policy optimization variants (GRPO and GSPO) and evaluate both functional correctness and behavioral diagnostics.
    In our experimental setting, RLVR improves \texttt{pass@1} on MBPP test by up to 13 percentage points under proposed combined reward configuration.
    However, we find that reward shaping can induce systematic behavioral shifts: using only static-analysis penalties may bias the policy toward shorter completions that reduce lint errors without reliably improving functional correctness.
    In contrast, combined rewards mitigate this degeneration and yield more stable trade-offs between correctness and style constraints.
    Overall, our results highlight that RLVR effectiveness for code generation is highly sensitive to reward design and optimization granularity, and that diagnostics beyond \texttt{pass@1}, including generation length, Ruff severity profiles, and execution error types are useful for identifying failure modes.
    \end{abstract}
	
	
	\maketitle
	
	\section{Introduction}
	\label{sec:intro}
	
	Code generation with large language models (LLMs) has become a widely used paradigm for automating routine programming tasks, assisted development, and rapid prototyping\cite{Jiang_2024}.
	Despite strong performance from instruction-tuned models, achieving high functional correctness remains challenging because training objectives (e.g., next-token prediction) are only indirectly aligned with execution behavior.
	A natural alternative is \emph{reinforcement learning with verifiable rewards} (RLVR) \cite{jiang2025coderlplus}, where rewards are computed from automatically checkable criteria such as compilation success, unit tests, or static analysis.
	
	In the code domain, unit tests provide a particularly compelling reward signal: they are objective, reproducible, and directly tied to correctness.
	However, RLVR is not only an ``accuracy improvement'' tool; it can also change generation behavior in non-trivial ways.
	Reward design and optimization choices can steer models toward undesirable equilibria, for example, reward hacking \cite{zhang2025surveyRL}.
    
    In this paper we test the following hypothesis using Mostly Basic Python Problems (MBPP) dataset: RLVR can substantially improve functional correctness, but its benefits are reward-sensitive; in particular, static-analysis-only shaping may degenerate toward shorter solutions, while a combined unit-test and static-analysis reward yields a more stable trade-off.
    
    We study RLVR in a controlled small-model setting where full training runs are feasible.
    We train LoRA adapters for two architectures (Qwen3-0.6B and Llama3.2-1B) and compare two group-based policy optimization variants, GRPO and GSPO, under multiple reward designs:
    (i) unit-test-only rewards, (ii) static-analysis-only shaping via Ruff, and (iii) combined rewards.
    By analyzing both correctness and behavioral diagnostics, we characterize when RLVR improves \texttt{pass@1} and when it instead shifts the model toward policies that primarily optimize proxy signals.

    Our contributions are as follows:
    \begin{itemize}
        \item we provide a controlled comparison of GRPO vs.\ GSPO for RLVR on MBPP using small models;
        \item we quantify the impact of different reward formulations on \texttt{pass@1} and on behavioral indicators (length, Ruff severity, and execution error types);
        \item  we document a reward-induced degeneration mode in our static-analysis-only setting and show that combined rewards mitigate this effect under the tested weights/normalization.
    \end{itemize}
    	
	\section{Related Work}
	\label{sec:related}

    \subsection{Code Generation Benchmarks}
	The evaluation of functional correctness in code generation is commonly performed using execution-based benchmarks such as HumanEval~\cite{humaneval2021} and Mostly Basic Python Problems (MBPP)~\cite{mbpp2021}. These benchmarks provide programming tasks paired with unit tests, allowing to track objective metrics such as \texttt{pass@k}.	
    For RLVR, these datasets are a natural fit because the evaluation signal (unit-test outcomes) matches the reward family used during training, enabling a direct assessment of whether reward optimization transfers to held-out tasks.
    Execution-based evaluation yields structured failure signals (e.g., syntax vs.\ runtime vs.\ assertion failures), and these error modes are informative for diagnosing reward-induced shifts.
    \subsection{RL for Code}
	Several works have demonstrated that incorporating execution feedback into reinforcement learning can improve functional correctness compared to purely supervised training.
	For example, CodeRL~\cite{le2022coderl} combines a pretrained code model with reinforcement learning driven by unit test outcomes and introduces critic-based mechanisms to mitigate reward sparsity.
    Subsequent studies have extended this line of work by designing richer feedback signals. For example, StepCoder introduces curriculum-style reinforcement learning with compiler/unit-test feedback \cite{dou2024stepcoder}. CodeRL+ further pushes this direction by introducing execution semantics alignment, adding variable-level execution trajectory signals so that optimization is guided by how the generated program behaves during execution \cite{jiang2025coderlplus}.
    However, RL with verifiable rewards is sensitive to reward shaping: auxiliary terms can introduce proxy objectives that are only loosely coupled to semantic correctness.
    Recent work has emphasized reward hacking as a practical concern in code-oriented RL environments and has proposed benchmarks and taxonomies for detecting reward exploits~\cite{deshpande2026benchmarkingrewardhackdetection}.

    \subsection{Group-Based Policy Optimization}
	Group Relative Policy Optimization (GRPO) was introduced in the DeepSeekMath work as a variant of Proximal Policy Optimization (PPO) that eliminates the separate critic using instead the baseline from group scores \cite{shao2024deepseekmath}.
	It has since been adopted in large-scale LLM reinforcement learning, including DeepSeek-R1, for optimizing reasoning and other verifiable tasks.
	
	A complementary formulation, Group Sequence Policy Optimization (GSPO), applies the group-relative objective at the sequence level rather than at the token level~\cite{gso2025group}.
	Both objectives are relevant for reinforcement learning with verifiable rewards (RLVR), where rewards are computed programmatically such as from unit tests.

	\section{Method Description}
	\label{sec:model}
    \subsection{Problem Setup}
    We study Python code generation from natural-language specifications. 
    Given a programming task described in text, the model must produce a Python implementation of the required function such that it satisfies a held-out set of unit tests.
    We use the Mostly Basic Python Problems dataset, where each instance consists of (i) a short problem statement and (ii) reference tests that specify the desired behavior.

    Formally, for each task $i$ we have a prompt $x_i$ (problem description, optionally augmented with a signature hint) and a candidate program $y_i$ generated by the model.
    Correctness is defined by execution: $y_i$ is considered correct if it compiles and passes all provided tests under the evaluation harness.
    A typical MBPP task asks for a small utility function, e.g
    \emph{``Write a python function to find the first repeated character in a given string''}
    
    The expected output is a self-contained Python function definition that can be executed by the test runner, for example:
    \begin{center}
        \texttt{assert first\_repeated\_char("abcabc") == "a"}
    \end{center}

	\subsection{Base models}
	We conduct experiments with two small LLMs:
	(i) \textbf{Qwen3-0.6B} \cite{qwen3technicalreport} and (ii) \textbf{Llama3.2-1B-Instruct} \cite{meta_llama32_1b_instruct}.
	Both models are used in their instruct variants.
	All training is performed with parameter-efficient adapters (LoRA) \cite{lora2021}, while the underlying base model parameters remain frozen.
	
	\subsection{Prompting}
	Each MBPP task provides a natural language problem statement and reference tests.
	During generation, the model is prompted with the problem description and the example of test case. It is important to note that we provide only one example of test case and not the entire list to avoid leaking potential solutions to the model. Example of the prompt is shown in Figure~\ref{fig:example_prompt}.
	We post-process model outputs by extracting Python code blocks and normalizing them into an executable module containing the target function.
    \begin{figure}[t]
    \centering
    \begin{minipage}{0.95\linewidth}
    \begin{lstlisting}[style=promptstyle]
    User: Please generate a python function for my problem. Do not provide any explanations, write only python code.

    Here is my problem:
    Write a python function to find the first repeated character in a given string.
    Test case example:
    assert first_repeated_char("abcabc") == "a"
    \end{lstlisting}
    \end{minipage}
    \caption{Example of generation prompt used in training process.}
    \label{fig:example_prompt}
    \end{figure}
    
    \subsection{Reward Design}
    We consider three reward formulations:
    \begin{itemize}
        \item Static-Analysis-Only Reward
        \item Unit-Test-Only Reward
        \item Combined Reward
    \end{itemize}
	\subsubsection{Static-Analysis-Only Reward}
	
	In the first RLVR configuration, rewards are derived exclusively from static analysis feedback produced by the \texttt{Ruff} linter \footnote{https://docs.astral.sh/ruff/}. 
	No unit test execution signal is used in this setting.
	The objective is to encourage syntactically correct and stylistically clean code while evaluating how such shaping influences downstream functional correctness. For each generated completion $y$, we invoke Ruff in JSON-output mode:
	\begin{center}
		\texttt{ruff check \texttt{-{}-}stdin-filename tmp.py \texttt{-{}-}output-format=json}
	\end{center}
	The generated source code is passed via standard input.
	Ruff returns a list of issues $\mathcal{E}(y)$, where each issue contains a rule identifier and diagnostic message.
	
	Each finding is mapped to a severity level:
	\[
	\text{severity}(e) \in \{\texttt{error}, \texttt{high}, \texttt{normal}, \texttt{low}\},
	\]
	using a predefined rule-to-severity mapping.
	If a rule identifier is not directly mapped, we attempt a fallback mapping using tokens extracted from the diagnostic message.
	Unmapped errors or warnings default to the \texttt{normal} severity level.
	Each severity level is associated with a fixed scalar penalty:
	\[
	w(\texttt{error}) = -10.0, \quad
	w(\texttt{high}) = -5.0, \quad
	w(\texttt{normal}) = -1.0, \quad
	w(\texttt{low}) = -0.2.
	\]
	
	Let $N_s(y)$ denote the number of findings of severity $s$ in completion $y$.
	The total Ruff-based reward is defined as
	\[
	R_{\text{ruff}}(y) =
	\sum_{s \in \{\texttt{error}, \texttt{high}, \texttt{normal}, \texttt{low}\}}
	w(s) \cdot N_s(y).
	\]
	
	If Ruff produces no issues (i.e., $\mathcal{E}(y)=\varnothing$) the reward is set to $0$.

    This severity-to-penalty mapping is heuristic rather than learned or tuned exhaustively. Its purpose is to encode an ordinal preference over Ruff findings, assigning larger penalties to more severe issues while keeping low-severity style violations weakly penalized. In this work, we use it as a simple and interpretable shaping signal rather than claiming it is the uniquely correct weighting scheme.

    \subsubsection{Length-Normalized Static-Analysis Reward}
    
    During the design of the static-analysis-only reward, we identified a potential failure mode: the model may reduce the total penalty simply by generating shorter and simpler code, without actually improving correctness.
    Because Ruff penalties are accumulated over the generated program, longer completions can receive larger negative rewards even when they represent reasonable solution attempts.
    
    To mitigate this length bias, we introduce a normalized variant of the static-analysis reward.
    Let
    \[
    \tilde{R}_{\text{ruff}}(y)=
    \sum_{s \in \{\texttt{error}, \texttt{high}, \texttt{normal}, \texttt{low}\}}
    w(s)\cdot N_s(y)
    \]
    denote the unnormalized Ruff penalty, and let $L(y)$ denote the number of generated lines in completion $y$.
    We define the normalized reward as
    \[
    R_{\text{ruff-norm}}(y)=
    \operatorname{clip}\!\left(
    \frac{\tilde{R}_{\text{ruff}}(y)}{\max(1,L(y))},
    -1,0
    \right).
    \]
    
    Thus, the normalized static-analysis reward remains penalty-only: completions with no Ruff findings receive reward $0$, while completions with more or more severe findings receive negative reward.
    Normalization reduces sensitivity to output length and discourages degenerate behavior in which the model minimizes penalties primarily by producing overly short code.
    
	\subsubsection{Unit-Test-Only Reward}
	\label{sec:unit_reward}
	In the second RLVR configuration, rewards are derived exclusively from functional correctness measured via unit test execution.
	No static-analysis penalties are used in this setting.
	The objective is to directly optimize behavioral correctness under the MBPP evaluation protocol.
	
	For each generated completion $y$, we extract the Python function implementation and evaluate it using the HumanEval-style execution harness.
	Each completion is paired with the corresponding list of MBPP tests, including both standard and challenge tests.
	
	Following prior work such as CodeRL~\cite{le2022coderl}, we employ a graded reward mapping that differentiates between distinct failure modes.
	The execution result for each completion is categorized into one of the following outcome classes:
	
	\begin{compactitem}
		\item \textbf{Passed}: all tests pass.
		\item \textbf{Assertion Error}: test assertions fail or execution times out.
		\item \textbf{Runtime Error}: runtime exceptions (e.g., \texttt{TypeError}, \texttt{IndexError}, \texttt{ValueError}, etc.).
		\item \textbf{Interpreter Error}: syntax or interpreter-level failures (e.g., \texttt{SyntaxError}, \texttt{NameError}, \texttt{IndentationError}).
		\item \textbf{Other}: uncategorized failures.
	\end{compactitem}
	
	Each outcome category is mapped to a scalar reward:
	
	\[
	R_{\text{unit}}(y) =
	\begin{cases}
		\phantom{-}1.0, & \text{if Passed}, \\
		-0.3, & \text{if Assertion Error}, \\
		-0.6, & \text{if Runtime Error}, \\
		-1.0, & \text{if Interpreter Error or otherwise}.
	\end{cases}
	\]
	
	This design introduces graded penalties rather than a purely binary reward.
	Interpreter-level failures receive the strongest penalty, reflecting complete invalidity of the program.
	Runtime errors receive intermediate penalties, while assertion failures incur milder penalties since the program structure is valid but functionally incorrect.
	
	\subsubsection{Combined Reward: Unit Tests + Normalized Ruff Penalty}
	
	In the third RLVR configuration, we combine execution-based unit test rewards with a normalized, penalty-only static analysis signal from Ruff.
	The unit test component remains identical to the unit-test-only setting (\ref{sec:unit_reward}), while the Ruff signal is normalized to reduce sensitivity to output length.
	In addition, we apply a small anti-shortcut penalty to discourage degenerate ``empty'' generations.
	
	Let $\mathcal{E}(y)$ denote the set of Ruff issues for completion $y$ and let $N_s(y)$ be the number of issues assigned to severity level $s \in \{\texttt{error},\texttt{high},\texttt{normal},\texttt{low}\}$ (using the same severity mapping as in the Ruff-only setting).
	We first compute the raw Ruff penalty as a weighted sum:
	\[
	R_{\text{raw}}(y) = \sum_{s} w(s)\,N_s(y),
	\]
	If Ruff produces no findings, $R_{\text{raw}}(y)=0$.
	If Ruff fails (e.g., invalid JSON output), we assign a mild penalty of $-0.5$.
	
	Because $R_{\text{raw}}(y)$ scales with the number of lines and may introduce a length bias, we normalize it by the number of lines:
	\[
	R_{\text{line}}(y)=\frac{R_{\text{raw}}(y)}{\max(1, L(y))},
	\]
	where $L(y)$ is the number of lines in $y$.
	
	Finally, we clip the normalized penalty into a bounded interval to prevent rare bursts of violations from dominating the reward:
	\[
	R_{\text{ruff}}(y) = \mathrm{clip}\big(R_{\text{line}}(y),\, [c_{\min}, 0]\big),
	\qquad c_{\min}=-1.0,
	\]
	so that $R_{\text{ruff}}(y)\in[-1,0]$.
	
	To discourage trivial completions that avoid both unit test failures and static analysis findings by producing almost no code, we apply a length-based penalty:
	\[
	R_{\text{short}}(y)=
	\begin{cases}
		-0.5, & \text{if } \mathrm{chars}(y) < 30 \ \text{or}\ L(y) < 3,\\
		\phantom{-}0, & \text{otherwise}.
	\end{cases}
	\]
	
	We combine unit test reward $R_{\text{unit}}(y)$ (\ref{sec:unit_reward}) with Ruff and anti-shortcut signals using a \emph{pass-gated} weighting scheme.
	The final reward is:
	\[
	R_{\text{comb}}(y)=
	\begin{cases}
		R_{\text{unit}}(y) + \lambda_{\text{pass}}\,R_{\text{ruff}}(y) + R_{\text{short}}(y),
		& \text{if } R_{\text{unit}}(y) = 1.0 \ (\text{Passed}), \\
		R_{\text{unit}}(y) + \lambda_{\text{fail}}\,R_{\text{ruff}}(y) + R_{\text{short}}(y),
		& \text{otherwise},
	\end{cases}
	\] where $\lambda_{\text{pass}}$, $\lambda_{\text{fail}}$ are specified constants introduced to regulate ``ruff`` component importance.
	
	This ``gate'' prevents Ruff penalties from dominating optimization on failing solutions (where functional correctness has not yet been achieved), while still encouraging cleaner code among already-correct solutions.
	
    \subsection{Group policy optimization}
    \label{sec:gpo}
    Given a prompt $x$, we sample a group of $G$ candidate completions $\{y_i\}_{i=1}^G$ from the current policy and compute scalar rewards $\{R_i\}$.
    We form a \emph{group-relative} advantage (TRL default is a simple group baseline; optionally normalized):
    \begin{equation}
    \label{eq:group-adv}
    \bar{R}=\frac{1}{G}\sum_{j=1}^G R_j,
    \qquad
    A_i = R_i - \bar{R}
    \quad
    \text{(optionally }A_i \leftarrow \tfrac{A_i}{\mathrm{std}(R)+\epsilon}\text{).}
    \end{equation}
    Let $y_i=(a_{i,1},\dots,a_{i,T_i})$ and denote the token-level likelihood ratio (importance weight) between the new policy $\pi_\theta$ and the sampling policy $\pi_{\theta_{\mathrm{old}}}$:
    \begin{equation}
    \label{eq:token-ratio}
    r_{i,t}(\theta)=\frac{\pi_\theta(a_{i,t}\mid x,a_{i,<t})}{\pi_{\theta_{\mathrm{old}}}(a_{i,t}\mid x,a_{i,<t})}.
    \end{equation}

    \paragraph{Token-level GRPO (TRL)} GRPO applies a PPO-style clipped policy-gradient objective at the \emph{token} level, using the same scalar group advantage $A_i$ for all tokens of a sampled completion:
    \begin{equation}
    \begin{aligned}
    \mathcal{L}_{\mathrm{GRPO}}(\theta)
    &= \frac{1}{G}\sum_{i=1}^G \frac{1}{T_i}\sum_{t=1}^{T_i}
    \min\!\Big(r_{i,t}A_i,\; \mathrm{clip}(r_{i,t},1-\varepsilon,1+\varepsilon)A_i\Big) \\
    &\quad - \beta\,\mathrm{KL}\!\big(\pi_\theta(\cdot\mid x)\,\|\,\pi_{\mathrm{ref}}(\cdot\mid x)\big).
    \end{aligned}
    \end{equation}
    where $\varepsilon$ is the clip range and $\beta$ controls KL regularization to a reference policy $\pi_{\mathrm{ref}}$.

    \paragraph{Sequence-level GSPO (GRPO variation in TRL):} GSPO uses the same group-relative advantage, but moves the importance weighting to the \emph{sequence} level by aggregating log-probability ratios across the entire completion:
    \begin{equation}
    \label{eq:seq-ratio}
    r_i(\theta)
    = \frac{\pi_\theta(y_i\mid x)}{\pi_{\theta_{\mathrm{old}}}(y_i\mid x)}
    = \exp\!\Big(\sum_{t=1}^{T_i}\log \pi_\theta(a_{i,t}\mid x,a_{i,<t})-\log \pi_{\theta_{\mathrm{old}}}(a_{i,t}\mid x,a_{i,<t})\Big).
    \end{equation}
    The clipped objective is then applied once per completion:
    \begin{equation}
    \begin{aligned}
    \mathcal{L}_{\mathrm{GSPO}}(\theta)
    &= \frac{1}{G}\sum_{i=1}^G
    \min\!\Big(r_i(\theta)A_i,\; \mathrm{clip}(r_i(\theta),1-\varepsilon,1+\varepsilon)A_i\Big) \\
    &\quad - \beta\,\mathrm{KL}\!\big(\pi_\theta(\cdot\mid x)\,\|\,\pi_{\mathrm{ref}}(\cdot\mid x)\big),
    \end{aligned}
    \end{equation}
    Compared to token-level GRPO, GSPO assigns a single importance weight to each sampled completion, which changes the credit assignment and can yield different length/entropy dynamics under the same reward.
	\section{Experiments}
    \label{sec:exp}
    \subsection{Dataset}
	\label{sec:data}

    We use the Mostly Basic Python Problems (MBPP) benchmark \cite{mbpp2021}, which consists of short Python programming tasks paired with reference solutions and unit tests. MBPP provides predefined splits; we use \textbf{MBPP train} for RLVR training and report primary results on the \textbf{MBPP test} split. 
    
    In addition, we perform a stricter external evaluation using EvalPlus benchmarks (MBPP-Plus and HumanEval-Plus) \cite{evalplus}, described in Section~\ref{sec:evalplus_eval}.

	We do not modify tasks or tests; all comparisons use the same evaluation harness.
    There are 974 tasks in total:
    \begin{itemize}
        \item 374 in train split,
        \item 500 in test split,
        \item 90 in validation split,
        \item 10 in ``prompt`` split (used for few-shot prompts)
    \end{itemize}
	\subsection{Metrics}
	We report \texttt{pass@1} on MBPP test, i.e., the fraction of tasks for which the first sampled solution passes all tests \cite{humaneval2021}.
	
	For failing tasks, we record the dominant failure mode from execution traces (e.g., \texttt{AssertionError}, \texttt{TypeError}, \texttt{NameError}, \texttt{SyntaxError}) and report category frequencies.
	
	We report the average number of Ruff findings per task, stratified by severity levels.
	
	We report mean generated tokens per run and study its correlation with \texttt{pass@1} and reward configurations.
	
	\subsection{Training Details}
	All RLVR training is performed using LoRA adapters \cite{lora2021}.
	We keep base weights frozen and optimize only adapter parameters.
	We use:
	\begin{compactitem}
		\item $\text{Number of generations per sample} = 8$,
        \item $\text{Batch Size (Prompts per Update) = 1}$,
        \item $\text{Gradient Accumulation Steps} = 1$,
		\item \texttt{learning rate} = \num{1e-5},
		\item \texttt{epochs} = 5,
		\item \texttt{sampling temperature} $= 0.9$ 
        \item \texttt{max tokens} $=768$,
		\item $Lora_{Rank} = 32$,
        \item $Lora_{\alpha} = 64$,
        \item KL coefficient $\beta = 0.0$ (no refernce model loaded),
        \item $\epsilon = 0.2$ (for clipping),
        \item $\lambda_{pass}=0.3$,
        \item $\lambda_{fail}=0.1$,
        \item $\text{Random State} = 42$
	\end{compactitem}
	For LoRa adapter we target several modules such as:
    \begin{itemize}
    \item Self-Attention projections:
        \begin{itemize}
            \item \texttt{q\_proj} - Query projection $W_Q$
            \item \texttt{k\_proj} - Key projection $W_K$
            \item \texttt{v\_proj} - Value projection $W_V$
            \item \texttt{o\_proj} - Attention output projection $W_O$
        \end{itemize}
        
        \item Feed-Forward (MLP) projections:
        \begin{itemize}
            \item \texttt{up\_proj} - Expansion projection $W_{\text{up}}$
            \item \texttt{gate\_proj} - Gating projection $W_{\text{gate}}$ 
            \item \texttt{down\_proj} - Contraction projection $W_{\text{down}}$
        \end{itemize}
    \end{itemize}
	We evaluate the following methods (matching the labels used in figures):
	\begin{compactitem}
		\item \textbf{normal}: base model inference with the same decoding parameters (no RLVR).
		\item \textbf{grpo-unit}, \textbf{gspo-unit}: RLVR with unit-test rewards only.
		\item \textbf{grpo-ruff}, \textbf{gspo-ruff}: RLVR with Ruff shaping only.
		\item \textbf{grpo-combined}, \textbf{gspo-combined}: combined reward.
	\end{compactitem}
		
	\subsection{Test evaluation}
	After fine-tuning with RLVR to evaluate the model's quality we make several runs. For each run we take MBPP test split, generate solutions for each task, using same methods as in training process we extract code, add test cases (both from basic test list and challenge test list) and evaluate the solution. We report \texttt{pass@1} for each run and provide $mean \pm SD$ for each method.
	We use decoding with sampling with temperature $T=0.6$ and max token count $max\_tokens=768$.

	Our primary baseline is the corresponding \textbf{normal} pretrained model for each architecture using the same prompt formatting and decoding settings.

    \subsection{Implementation}
    Training and evaluation scripts were done with Python 3.12.12, using Transformers v4.57.6 \cite{hf_transformers}, TRL 0.24 \cite{hf_trl}, Unsloth 2026.1.4 \cite{unsloth} with Pytorch 2.10. For code and test execution we utilize OpenAI human-eval python module \footnote{\url{https://github.com/openai/human-eval}}.
    Static analysis feedback was provided by using Ruff 0.14.11 \footnote{\url{https://docs.astral.sh/ruff}} with default ruleset (Flake8's \textit{F} rules, along with a subset of the \textit{E} rules).
    
    All training runs were done on rented GPU-server with Nvidia RTX 3090 (24GB VRAM), while test evaluations were performed on PC with Nvidia RTX 5060Ti (16GB VRAM).
    
    \subsection{EvalPlus evaluation}
    \label{sec:evalplus_eval}
    To complement the main MBPP test evaluation, we additionally evaluate selected reward configurations using EvalPlus \cite{evalplus} on MBPP+ and HumanEval+, which provide a stricter hidden-test benchmark for functional correctness. For this evaluation, we retrain each reward configuration with two random seeds (101 and 202) and report per-seed pass@1 results. Inference is performed with an OpenAI-compatible vLLM backend (v0.17.0) \cite{vllm} using greedy decoding. We evaluate on both the base and the additional hidden tests provided by EvalPlus.
	\section{Results and Discussion}
	\label{sec:results}
	
	\subsection{Overall Performance}
	For each combination of model and method we ran 10 test evaluations.
	Table~\ref{tab:main_results} reports mean \texttt{pass@1} on MBPP test set and p-value. We compute the p-value with a paired randomization (sign-flip) test over tasks: we repeatedly flip the sign of each task’s method baseline score difference at random to simulate the null hypothesis of ``no effect`` and report the fraction of simulations where the resulting mean difference is at least as extreme as the observed one.
	We observe consistent gains from RLVR with unit-test only and combined rewards across both model families, with the best combined-reward configuration of Qwen3-0.6B model using Combined Reward improving \texttt{pass@1} by up to 13 percent points over the baseline.
	Across the reported settings, neither GRPO nor GSPO consistently dominates the other. A plausible explanation is that MBPP and HumanEval consist of relatively short, self-contained tasks with short prompts and completions, so the benefit of sequence-level optimization is less pronounced than it might be in longer-horizon settings.
	\begin{table}[!tbh]
		\centering
		\caption{MBPP test pass@1 (mean $\pm$ SD over runs). $\Delta$ indicates absolute improvement over the corresponding normal baseline.}
		\label{tab:main_results}
		\footnotesize
		\begin{tabular}{l|cccc}
			\toprule
			\textbf{Model} & \textbf{Method} & \textbf{pass@1} & \textbf{$\Delta$ vs normal} & p-value \\
			\midrule
			\multirow{8}{*}{Qwen3-0.6B}
			& gspo-combined & \textbf{0.417 $\pm$ 0.016} & +0.144 & 5.0e-6 \\ 
			& grpo-combined & 0.415 $\pm$ 0.010 & +0.142 & 5.0e-6 \\
			& gspo-unit & 0.412 $\pm$ 0.010 & +0.139 & 5.0e-6 \\
			& grpo-unit & 0.388 $\pm$ 0.012 & +0.115 & 5.0e-6 \\
			& normal & 0.273 $\pm$ 0.007 & -- & -- \\
            & grpo-ruff-normalized & 0.253 $\pm$ 0.006 & -0.020 & 5.0e-6 \\
			& grpo-ruff & 0.252 $\pm$ 0.009 & -0.021 & 2.6e-2 \\
            & gspo-ruff-normalized & 0.249 $\pm$ 0.008 & -0.024 & 5.0e-6 \\
			& gspo-ruff & 0.232 $\pm$ 0.008 & -0.041 & 3.5e-4 \\
			\midrule
			\multirow{7}{*}{Llama3.2-1B}
			& grpo-combined & \textbf{0.389 $\pm$ 0.008} & +0.040 & 9.5e-4 \\
			& gspo-combined & 0.383 $\pm$ 0.009 & +0.034 & 4.9e-3 \\
			& grpo-unit & 0.379 $\pm$ 0.010 & +0.030 & 1.8e-3\\
			& gspo-unit & 0.375 $\pm$ 0.014 & +0.026 & 4.7e-3 \\
			& normal & 0.349 $\pm$ 0.010 & -- & -- \\
			& grpo-ruff & 0.313 $\pm$ 0.011 & -0.036 & 4.5e-4 \\
			& gspo-ruff & 0.312 $\pm$ 0.010 & -0.037 & 3.5e-5 \\
            & grpo-ruff-normalized & 0.294 $\pm$ 0.010 & -0.055 & 5.0e-6 \\
            & gspo-ruff-normalized & 0.287 $\pm$ 0.011 & -0.062 & 5.0e-6 \\
			\bottomrule
		\end{tabular}
	\end{table}

	\subsection{Error Analysis}
\begin{figure}[!tbh]
		\centering
		\includegraphics[width=0.98\textwidth]{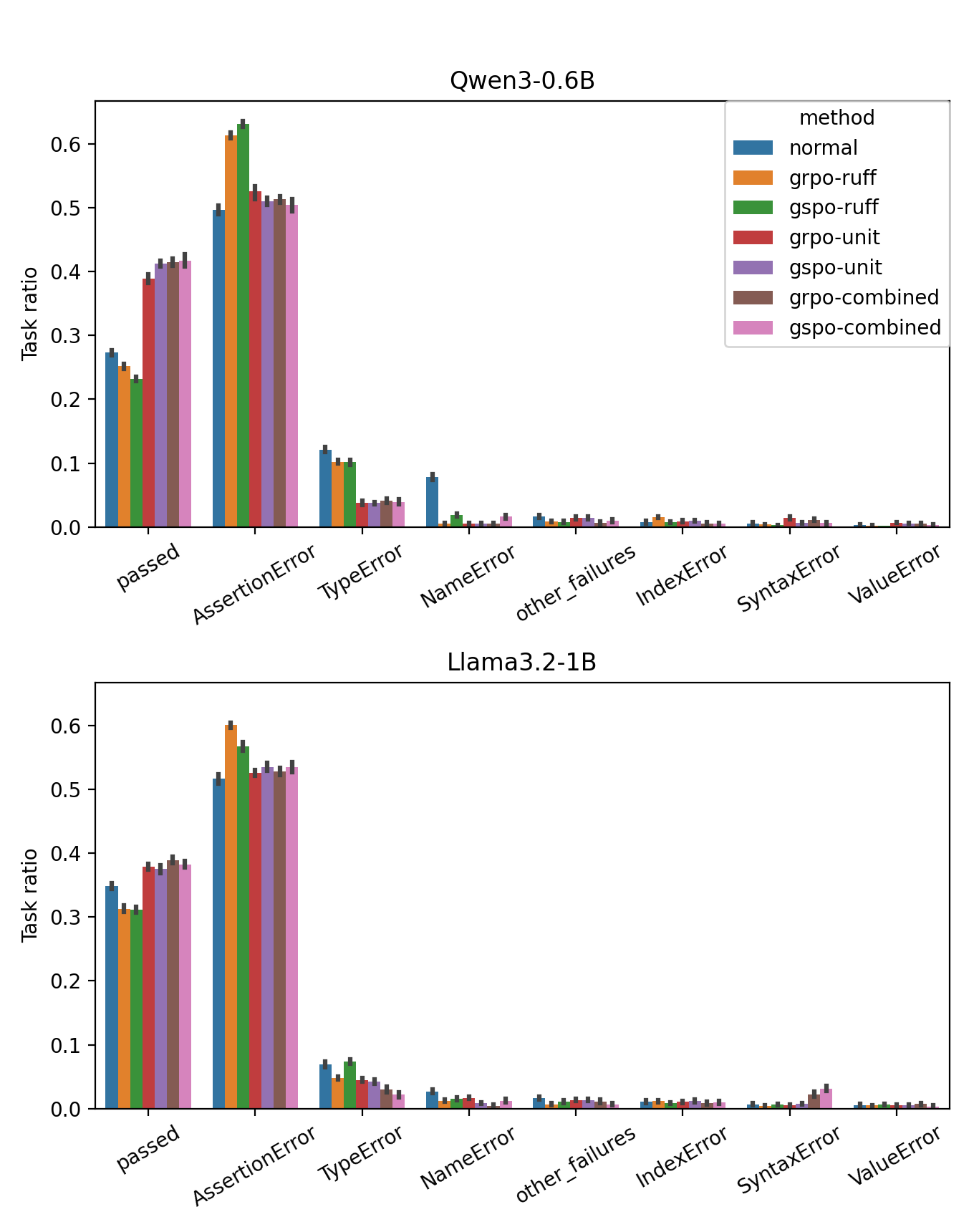}
		\caption{Execution outcome distribution on MBPP test (mean over runs).}
		\label{fig:exec_errors}
	\end{figure}

	Figure~\ref{fig:exec_errors} shows the distribution of execution outcomes.
	For combined and unit-test only we can see reduction of AssertionErrors and an increase in the fraction of passed solutions.
	We also observe method-specific shifts among runtime and interpreter errors (e.g., \texttt{NameError}, \texttt{TypeError}), suggesting that different objectives may prioritize distinct correctness improvements (semantic vs. syntactic robustness).
	
	\subsection{Static Analysis Effects}
\begin{figure}[]
		\centering
		\includegraphics[width=0.98\textwidth]{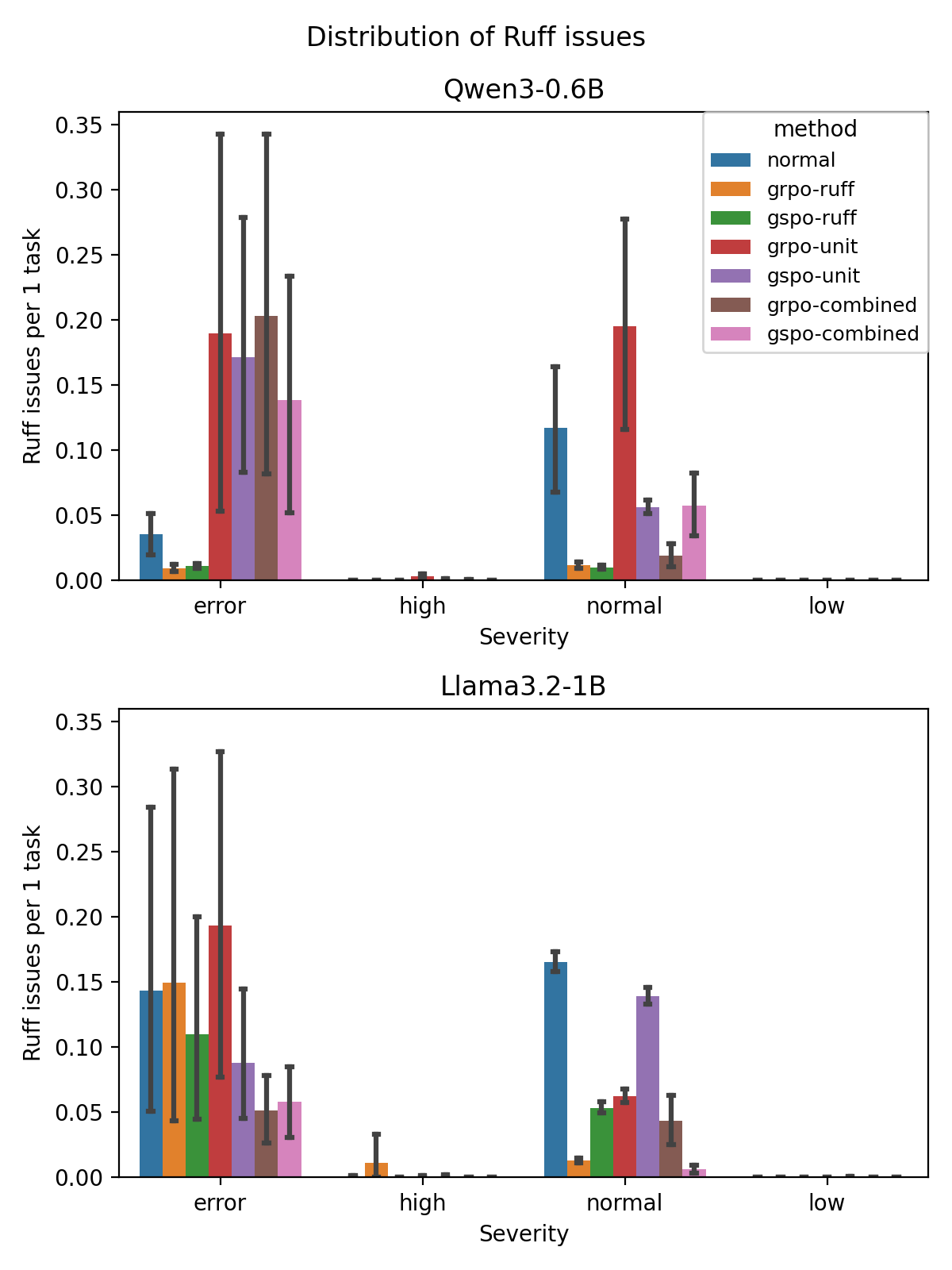}
		\caption{Ruff finding severity distribution (average findings per task).}
		\label{fig:ruff_severity}
	\end{figure}

	Figure~\ref{fig:ruff_severity} reports Ruff findings by severity.
	Static-analysis shaping substantially changes the distribution of findings, but these improvements do not always translate proportionally into higher \texttt{pass@1}.
	This highlights a key RLVR tradeoff: optimizing a proxy reward (style/quality signals) can steer generation away from test-passing correctness if not calibrated carefully.
	
	\subsection{Length Degeneration}
\begin{figure}[]
		\centering
		\includegraphics[width=0.98\textwidth]{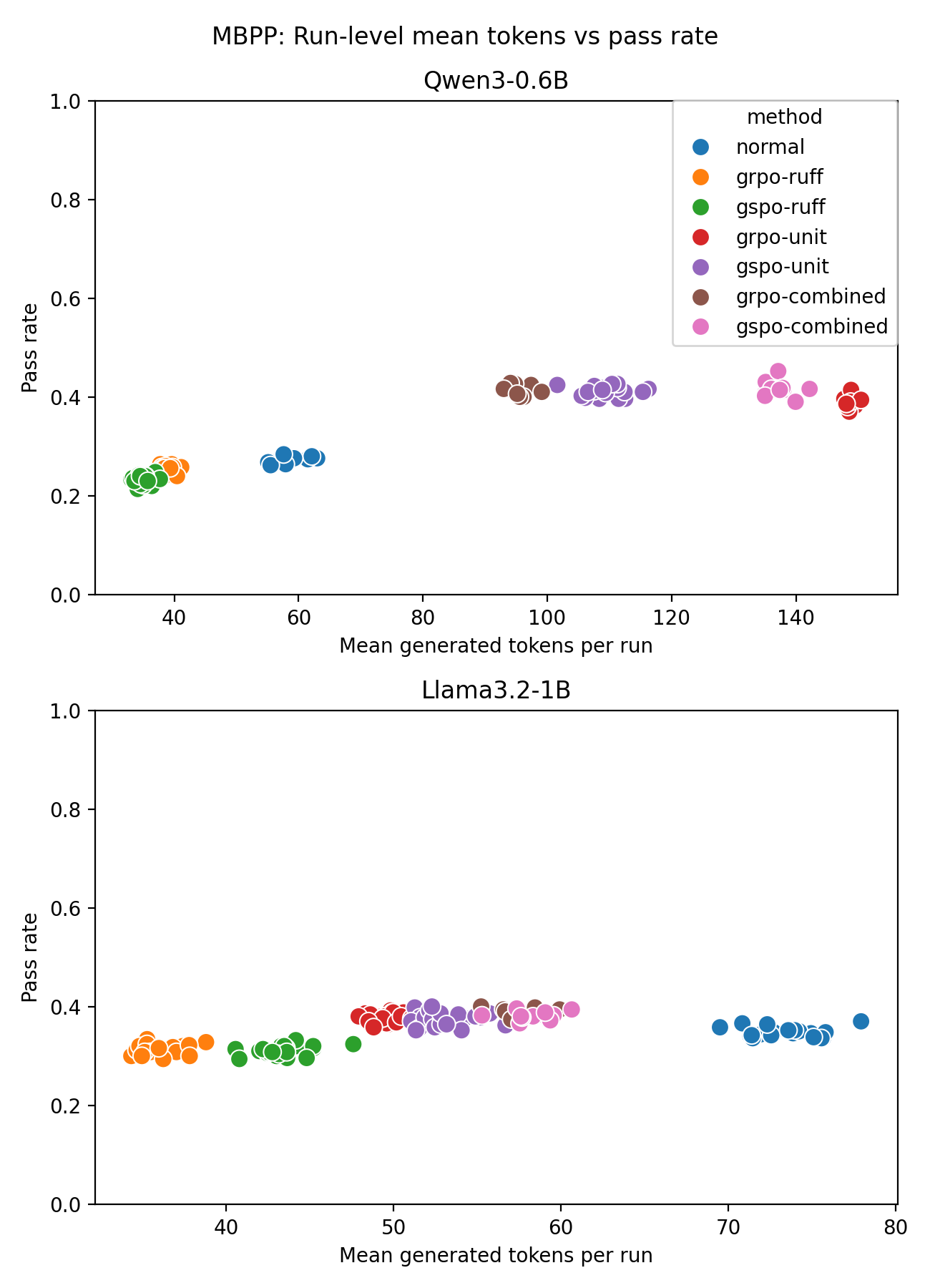}
		\caption{Run-level mean generated tokens vs MBPP pass rate.}
		\label{fig:tokens_vs_pass}
	\end{figure}

	Figure~\ref{fig:tokens_vs_pass} plots mean number of generated tokens vs pass rate across runs.
	We can observe that Ruff-dominant configuration tend to produce shorter output while having lower pass rate. This static-analysis only configuration skews model into reward-driven degeneration which incentives simple code that reduces style violations without improving functional correctness.
	While unit-test configuration tend to produce longer outputs with bigger pass rates.
	These findings motivate reporting behavioral diagnostics alongside \texttt{pass@1} for RLVR-trained code models.
	
    \subsection{External Validation with EvalPlus}
    So far, the previous results were based on the MBPP test split and its associated diagnostics. To assess whether these patterns transfer to a stricter hidden-test setting, we additionally evaluate independently retrained adapters on EvalPlus benchmarks (MBPP+ and HumanEval+). Table~\ref{tab:evalplus_results} reports per-seed results for both model families.
    
    The overall pattern remains similar to the main MBPP analysis. Static-analysis-only methods remain unstable and often weak, while the unit-test-only and combined reward families are generally stronger. However, under stricter evaluation, the combined reward is competitive in some settings but does not provide a consistent improvement over the unit-test-only reward across both models and both benchmarks. These results suggest that the benefits of reward shaping are sensitive to evaluation protocol and do not always transfer uniformly to stronger hidden-test settings.

    \begin{table}[!tbh]
        \centering
        \caption{Per-seed EvalPlus results for independently retrained adapters.}
        \label{tab:evalplus_results}
        \scriptsize
        \setlength{\tabcolsep}{4pt}
        \begin{tabular}{lcccccccc}
            \toprule
            & \multicolumn{4}{c}{Qwen3-0.6B} & \multicolumn{4}{c}{Llama-3.2-1B} \\
            \cmidrule(lr){2-5} \cmidrule(lr){6-9}
            & \multicolumn{2}{c}{MBPP+} & \multicolumn{2}{c}{HumanEval+} & \multicolumn{2}{c}{MBPP+} & \multicolumn{2}{c}{HumanEval+} \\
            \cmidrule(lr){2-3} \cmidrule(lr){4-5} \cmidrule(lr){6-7} \cmidrule(lr){8-9}
            Method & 101 & 202 & 101 & 202 & 101 & 202 & 101 & 202 \\
            \midrule
            Base        & 0.405 & 0.405 & 0.384 & 0.384 & 0.407 & 0.407 & 0.311 & 0.311 \\
            \midrule
            Ruff GRPO   & 0.405 & 0.267 & 0.256 & 0.244 & 0.429 & 0.399 & 0.299 & 0.293 \\
            Ruff GSPO   & 0.265 & 0.344 & 0.244 & 0.268 & 0.429 & 0.302 & 0.287 & 0.305 \\
            \midrule
            Ruff-norm GRPO & 0.357 & 0.302 & 0.299 & 0.268 & 0.386 & 0.399 & 0.341 & 0.317 \\
            Ruff-norm GSPO & 0.399 & 0.312 & 0.280 & 0.274 & 0.376 & 0.407 & 0.323 & 0.329 \\
            \midrule
            Unit-tests GRPO     & 0.508 & 0.526 & 0.396 & 0.396 & 0.434 & 0.437 & 0.341 & 0.354 \\
            Unit-tests GSPO     & 0.460 & \textbf{0.537} & 0.323 & 0.366 & 0.450 & 0.447 & 0.335 & 0.335 \\
            \midrule
            Combined GRPO  & 0.484 & 0.503 & \textbf{0.402} & 0.396 & 0.442 & 0.452 & 0.360 & 0.348 \\
            Combined GSPO  & 0.526 & 0.505 & 0.372 & 0.396 & \textbf{0.471} & 0.450 & \textbf{0.439} & 0.348 \\
            \bottomrule
        \end{tabular}
    \end{table}

    \section{Limitations}
    Currently all the experiments were done with test split of MBPP dataset and EvalPlus benchmark. These benchmarks consist of short, mostly self-contained Python programming tasks with relatively small test suites. 
    As a result, conclusions may not transfer to long-horizon code generation settings 

    Our reward formulations include several heuristic design decisions:
    \begin{itemize}
        \item a graded unit-test reward mapping,
        \item a Ruff-derived penalty,
        \item a line-normalized and clipped Ruff reward in the combined setting
    \end{itemize}
    These choices were motivated by stability and practical considerations, but they introduce additional hyperparameters (e.g., clipping threshold, weighting ($\lambda$) coefficients, length thresholds) that may affect outcomes.
    We did not perform exhaustive tests over these parameters, and alternative normalizations or weight schedules could yield different trade-offs.

    We evaluate two small models (Qwen3-0.6B and Llama3.2-1B), which still leaves open whether the same reward-shaping effects hold for larger models or for models trained with different instruction-tuning schemes.
	\section{Conclusion}
	\label{sec:conc}
    We presented a study of reinforcement learning with verifiable rewards (RLVR) for Python code generation on MBPP using two small language models, Qwen3-0.6B and Llama3.2-1B.
    Across both architectures, RLVR improves functional correctness, with the best configuration achieving up to +13 percent points in \texttt{pass@1} over the corresponding baseline.
    Regarding the optimization variants, we do not observe a universal winner between token-level GRPO and sequence-level GSPO. Both variants are capable of improving performance under suitable reward design, and neither is uniformly dominant across any tests. A more decisive comparison between GRPO and GSPO may require longer-horizon code-generation benchmarks, where sequence-level credit assignment is expected to matter more strongly than in short benchmark tasks such as MBPP and HumanEval.
    Our analysis also shows that reward design induce behavioral shifts. 
    First, unit-test-driven RLVR substantially changes the failure profile, we can observe not only reduction in Assertion Errors (failing unit tests), but also with Type and Name Errors.
    Second, static-analysis rewards affects code quality diagnostics but does not necessarily translate into higher \texttt{pass@1}: Ruff-optimized runs shift severity distributions toward fewer ``error``-level severity issues, yet can underperform in functional correctness when the static penalty dominates the learning signal.
    Third, we find that static-analysis penalties interact strongly with output length.
    According to our tests, Ruff-heavy objectives tend to produce shorter generations, whereas combined rewards shift the model toward longer completions that correlate with higher \texttt{pass@1}.
    Overall, the results emphasize that RLVR for code should be evaluated with diagnostics beyond \texttt{pass@1}, including execution error categories, static-analysis profiles, and length statistics.
    
    Future work includes more extensive reward-weight sweeps, training-dynamics analysis (KL/reward curves), systematic sensitivity analysis of the Ruff severity weights, including alternative scaling schemes and severity groupings and extending the approach to additional datasets and languages.
	\providecommand{\bysame}{\leavevmode\hbox to3em{\hrulefill}\thinspace}
	\providecommand{\MR}{\relax\ifhmode\unskip\space\fi MR }
	\providecommand{\MRhref}[2]{\href{http://www.ams.org/mathscinet-getitem?mr=#1}{#2}}
	\providecommand{\href}[2]{#2}

\end{document}